\documentclass{article}
\usepackage{amsmath}
\usepackage{amsfonts}
\usepackage{latexsym}
\usepackage{amssymb}
\usepackage{eepic}
\usepackage{epic}
\usepackage{epsfig}
\usepackage{multirow}
\usepackage{color}

\numberwithin{equation}{section}

\begin{document}

\title{Linear Augmented Slater-Type Orbital Method for Free Standing Clusters}

\author{K. S. Kang ${}^1$, J. W. Davenport${}^{1,2}$, J. Glimm${}^{1,3}$, \\
D. E. Keyes${}^4$, and M. McGuigan${}^1$\\
{}\\
${}^1$ Computational Science Center, \\
Brookhaven National Laboratory\\
${}^2 $Center for Functional Nanomaterials,\\
Brookhaven National Laboratory\\
${}^3$ Department of Applied Mathematics,\\
Stony Brook University\\
${}^4$ Department of Applied Physics and Applied Mathematics,\\
Columbia University
}
\date{}
\maketitle

\begin{abstract}
We have developed a  Scalable Linear Augmented Slater-Type Orbital
(LASTO) method for electronic-structure calculations on free-standing atomic clusters. As with other linear methods we solve the
Schr\"odinger equation using a mixed basis set consisting of
numerical functions inside atom-centered spheres and matched onto
tail functions outside. The tail functions are Slater-type orbitals,
which are localized, exponentially decaying functions. To solve the
Poisson equation between spheres, we use a finite difference method
replacing the rapidly varying charge density inside the spheres with
a smoothed density with the same multipole moments. We use multigrid
techniques on the mesh, which yield the Coulomb potential on the spheres and in turn defines
the potential inside via a Dirichlet problem. To solve the linear
eigen-problem, we use ScaLAPACK, a well-developed package to solve
large eigensystems with dense matrices. We have tested the method
on small clusters of palladium. 
\end{abstract}

%\maketitle
%\begin{keywords}
%multigrid method
%\end{keywords}

\section{Introduction}

There are many ways to solve the coupled Schrodinger and Poisson
equations required for density functional theory.  They generally
fall into two classes: (a) finite cluster calculations using basis
sets, such as Gaussians, or (b) periodic crystal calculations, as for
example those which use a plane wave basis set. Within either class the nuclear attraction may be replaced by a pseudo or effective core potential.

The ability to fabricate nanoscale clusters with tens of thousands
of atoms has driven a renewed interest in electronic structure
methods capable of reaching this size.  In our research we have
opted for the finite cluster approach, since many of the systems of
interest fall into this category.  However, the standard techniques
using Gaussians often require extremely large basis sets and are
difficult to apply to heavy atoms.  Methods for treating crystals
which derive from the augmented plane wave (APW) method are known to be extremely accurate
and suitable for all atoms in the periodic table but make essential
use of Fourier decompositions in a way that does not scale well to
massively parallel machines.  In addition, they make use of
supercells, which may not describe finite systems accurately.

Some years ago we developed an augmented basis set method which uses
linearized solutions of the Schrodinger (or Dirac) equation inside
atom centered spheres, and Slater-type-orbitals in the region
between spheres \cite{daven, daven1, ferna}. It was based on Anderson's linear muffin tin orbital method (LMTO) \cite{Anderson} but used Slater type orbitals as “tail” functions in place of Anderson's Spherical Bessel functions. While the method was
formulated in real space, it was more cleanly coded in reciprocal
space, which for crystals with small unit cells was equally
efficient.  

We return here to the real space formulation and apply
it to free standing clusters. There are a number of other real space formulations of density functional theory (for a review see \cite{Beck}). Most of these solve the Schr\"odinger equation directly on a mesh, as for example Chelikowski $et$ $al$ \cite{Chelikowski,
Li}.

An essential difference is the method for solving the Poisson
equation.  The standard method in crystals \cite{Weinert, Mattheiss} is
to replace the charge density inside the spheres by a smooth
pseudo-density which has the same multipole moments as the true
density (including the nuclear charge) and is represented by a
Fourier series.  Here we use the same idea, except we represent the
pseudo-density on a (nonuniform) grid to enhance resolution adaptively, not globally.  In addition, we use the
pseudo-density only for the spherical (monopole) portion of the
density, as the nonspherical terms can be included directly on the
grid.

In either case, one finds the solution of the Poisson equation
outside the spheres and interpolates onto the sphere boundaries to
define a Dirichlet problem for the potential inside which can be
solved using the true density.

\section{Methods}

In this section, we consider the Kohn-Sham approach in real space
for electronic structure calculations.

 We consider the Schr\"odinger
equation
\begin{equation}\label{sch1}
\left[ -\frac{\nabla^2}{2} + V(\rho(\bold x))\right]
\varphi_k(\bold x) = \epsilon_k\varphi_k(\bold x),
\end{equation}
where
\begin{equation}
\rho(\bold x) = \sum_{\epsilon_k < E_{\mbox{\scriptsize Fermi}}}
|\varphi_k(\bold x)|^2 +\rho_{\mbox{\scriptsize core}}(V(\rho(\bold x)))
\end{equation}
is the electronic charge density and
\begin{equation}  V(\rho(\bold x))  = V_{\mbox{\scriptsize Ext}}(\bold x) +
V_{\mbox{\scriptsize Hartree}}(\rho(\bold x)) +
V_{\mbox{\scriptsize xc}}(\rho(\bold x)).
\end{equation}
is the potential, which consists of the external potential
$V_{\mbox{\scriptsize Ext}}$, the Hartree potential
$V_{\mbox{\scriptsize Hartree}}$, and the exchange and correlation
potential $V_{\mbox{\scriptsize xc}}$.  We use atomic units where lengths are measured in units of the Bohr radius $a_0 \hbar^2/me^2 = 0.529$ $\AA$ and energies in Hartrees $e^2/a_0 = 27.212$ $eV$. The external potential
$V_{\mbox{\scriptsize Ext}}$ is typically a sum of nuclear
potentials centered at the atomic positions. The Hartree potential
$V_{\mbox{\scriptsize Hartree}}$ can be obtained by solving the
Poisson equation
\begin{equation} \label{poie1}
 \nabla^2V_{\mbox{\scriptsize Hartree}}(\bold x) = 4\pi \rho(\bold
x) \end{equation} with
\begin{equation}
 \lim_{|\bold x| \to \infty} V_{\mbox{\scriptsize Hartree}}(\bold x)
 = -\lim_{|\bold x| \to \infty} V_{\mbox{\scriptsize Ext}}(\bold x),
\end{equation}
where $$\nabla^2V_{\mbox{\scriptsize Ext}}(\bold x) = 4\pi\sum_{j}
Z_j\delta_{\bold x = C_j}(\bold x), $$ where $C_j$'s are the center
of atoms and $Z_j$'s are the nuclear charge. We define the Coulomb
potential $V_C(\bold x) = V_{\mbox{\scriptsize Ext}}(\bold x) +
V_{\mbox{\scriptsize Hartree}}(\rho(\bold x))$ and solve the
Poisson equation
\begin{equation} \label{poie1}
 \nabla^2V_C(\bold x) =  4\pi (\rho(\bold
x) - \sum_{j}Z_j\delta_{\bold x = C_j}(\bold x) ) \end{equation}
with
\begin{equation}
 \lim_{|\bold x| \to \infty} V_C(\bold x) = 0.
\end{equation}

 The exchange and correlation potential
$V_{\mbox{\scriptsize xc}}$ is formally defined through the
functional derivative of the exchange and correlation energy,
\begin{equation}
V_{\mbox{\scriptsize xc}} = \frac{\delta E_{\mbox{\scriptsize
xc}}}{\delta \rho}\Big|_{\bold x}.
\end{equation}
In principle, only the density functionals for the exchange and
correlation term remain to be approximated. Many approximations for
the exchange-correlation functionals have been developed. We use local
spin density (LSD) of the Hedin-Lundqvist form \cite{martin} and
generalized gradient approximations (GGA) of Perdew, Burke, Wang
and Ernzerhof, (PBE-GGA)\cite{pbw96}. The GGA
approximation is generally considered to be more accurate and we
confirm this view in our work.

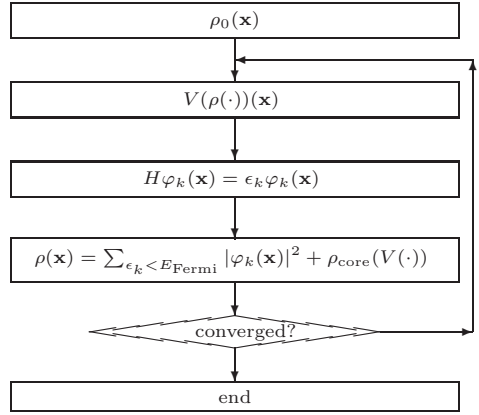
\begin{figure}
\begin{center}
\setlength{\unitlength}{1pt} \scriptsize
\begin{picture}(200,160)(0,-10)
\put(10,140){\framebox[170pt][c]{$\rho_0(\bold x)$}}
\put(10,110){\framebox[170pt][c]{$V(\rho(\cdot))(\bold x)$ }}
\put(10,80){\framebox[170pt][c]{$H\varphi_k(\bold x) =
\epsilon_k\varphi_k(\bold x)$ }}
\put(10,50){\framebox[170pt][c]{$\rho(\bold x) = \sum_{\epsilon_k <
E_{\mbox{\tiny Fermi}}} |\varphi_k(\bold x)|^2 +\rho_{\mbox{\tiny
core}}(V(\cdot))$ }} \drawline(95,30)(40,24)(95,18)(150,24)(95,30)
\put(80,22){\mbox{converged?}}
\put(10,-3){\framebox[170pt][c]{\mbox{end}}}
\put(95,135){\vector(0,-1){16}}
\put(95,105){\vector(0,-1){16}}\put(95,75){\vector(0,-1){16}}
\put(95,43){\vector(0,-1){13}}\put(95,18){\vector(0,-1){13}}
\put(150,24){\vector(1,0){35}}\put(185,127){\vector(-1,0){90}}
\put(185,24){\vector(0,1){103}}
\end{picture}
\caption{Flow chart for the solving Schr\"odinger equation with DFT
$*$}\label{chart}
\end{center}
\end{figure}

 Due to the functional dependence of $V$ on the density, these equations
form a set of nonlinear coupled equations. The standard solution
procedure is to iterate on the solution of the linear subsystems
until self-consistency is achieved.
We show the basic flow-chart to solve the Schr\"odinger equation
with DFT in Fig. \ref{chart}.

To solve the Schr\"odinger equation, we use the linear
augmented-Slater-type-orbital (LASTO) basis set which has been
described in detail in \cite{daven,daven1,ferna}. Here, we summarize
results for the LASTO basis set for numerical implementation.

\begin{figure}
\begin{center}
\setlength{\unitlength}{2pt}
\begin{picture}(100,100)(0,0)
\put(30,30){\makebox(0,0){$\bullet$}} \put(30,30){\circle{33}}
\drawline(30,30)(46.5,30) \drawline(70,70)(86.5,70)
\put(35,33){\makebox{$R_1$}} \put(75,73){\makebox{$R_2$}}
\put(15,45){\makebox{$S_1$}} \put(55,85){\makebox{$S_2$}}
\put(70,70){\makebox(0,0){$\bullet$}}\put(70,70){\circle{33}}
\drawline(5,5)(95,5)(95,95)(5,95)(5,5)
\end{picture}
\caption{Simplified domain in 2D.}\label{figure2}
\end{center}\end{figure}
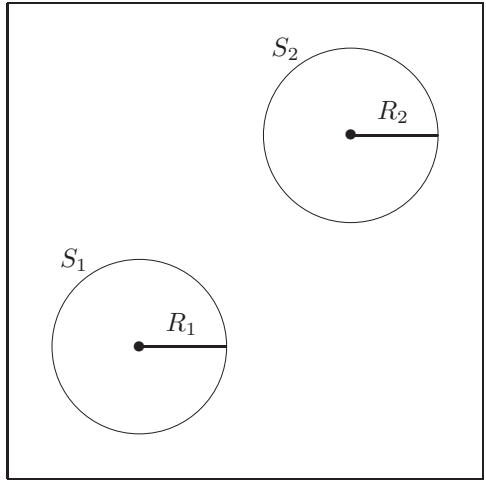

We introduce a sphere $S_j$ around the $j$th atom (Fig.
\ref{figure2}). We use different functional forms for the basis
functions, $\phi_{inlm}$ which is defined with reference atom $i$,
inside the $S_j$ and the region outside all $S_j$, called the
interstitial region;
\begin{equation}\label{sto}
\phi_{inlm}(\bold x) = \begin{cases} r_i^{n-1}e^{-\zeta r_i}
y_{lm}(\hat{r}_i), \quad \mbox{in the interstitial
region},\\
\sum_{j\lambda\mu} [\beta_{inlm,j\lambda\mu}g_{j\lambda}(r_j)
+\alpha_{inlm,j\lambda\mu}\dot{g}_{j\lambda}(r_j)]
y_{j\lambda\mu}(\hat{r}_j), \quad\mbox{ inside $S_j$},
\end{cases}
\end{equation}
where $r_j$ and $\hat{r}_j = (\theta,\phi)$ are the spherical
coordinates for $\bold x = (x,y,z) = (r \sin\theta\cos \phi, r
\sin\theta \sin\phi ,r\cos \theta)$ with respect to the position $X_j$
of atom $j$, $S_j$ is a sphere centered at the atom with radius $R_j$,
and the $y_{lm}$ are real spherical harmonics.

In (\ref{sto}), the $g_{j\lambda}$ are numerical solutions of the
scalar relativistic radial Dirac equation and the
$\dot{g}_{j\lambda}$ are their energy derivatives. They satisfy the
radial equations
\begin{equation} \label{pra1}
h_r g_{j\lambda} = e_j g_{j\lambda}\end{equation}
 and
 \begin{equation} \label{pra2}
h_r\dot{g}_{j\lambda} = e_j \dot{g}_{j\lambda} +
g_{j\lambda},\end{equation} where $h_r$ is the scalar relativistic
radial Hamiltonian,
$$ h_r = -\frac{\hbar^2}{2M}\left[\frac{d^2}{dr^2}
+\frac{2}{r}\frac{d}{dr} -\frac{\lambda(\lambda+1)}{r^2}\right]
-\frac{\hbar^2}{2M} \frac{1}{2Mc^2}\frac{dV_0}{dr}\frac{d}{dr} +
V_0,$$
$$ M = m\left[1+\frac{e_j -V_0}{2mc^2}\right],$$
 and $V_0$ is the spherical average of the potential.
The $g$'s are normalized within the spheres,
$$\int_0^{R_S}r^2 g_{j\lambda}^2(r) dr = 1,$$
and $g_{j\lambda}$ and $\dot{g}_{j\lambda}$ are orthogonal.

The $\beta$'s and $\alpha$'s are chosen by matching the interior and
the exterior functions and their derivatives on the boundaries of
the spheres, i.e.,
\begin{align*}
\phi_{inlm}(\bold x) &=
\sum_{\lambda\mu}[\beta_{inlm,j\lambda\mu}g_{j\lambda}(r_j) +
\alpha_{inlm,j\lambda\mu} \dot{g}_{j\lambda}(r_j)]
y_{\lambda\mu}(\hat{r}_j) \\
& =r_i^{n-1}e^{-\zeta r_i} y_{lm}(\hat{r}_i)
=\sum_{\lambda\mu}C_{ilnm,j\lambda\mu}(r_j)
y_{\lambda\mu}(\hat{r}_j)
,\\
 \frac{d }{d r_j} \phi_{inlm}(\bold x) &=
\sum_{\lambda\mu}[\beta_{inlm,j\lambda\mu}g'_{j\lambda}(r_j) +
\alpha_{inlm,j\lambda\mu} \dot{g}'_{j\lambda}(r_j)]
y_{\lambda\mu}(\hat{r}_j) \\
& =\frac{d}{d r_j} r_i^{n-1}e^{-\zeta r_i} y_{lm}(\hat{r}_i)
=\sum_{\lambda\mu}C'_{ilnm,j\lambda\mu}(r_j)
y_{\lambda\mu}(\hat{r}_j)
\end{align*}
for $r_j = R_j$. Computation of the $C_{ilnm,j\lambda\mu}$ and
$C'_{inlm,j\lambda\mu}$ require the expansion of an STO about a
site other than the one on which it is centered. This problem has
been considered by many authors \cite{daven}. Here we only summarize
the results:
\begin{align}\label{pco1}
C_{inlm,j\lambda\mu} & =  \begin{cases} 4\pi \sum_{l''}
I_R(lm,l'm',l''m'')
V_{ll'l''}^n(r,R_{ji})y_{l''m''}(\hat{R}_{ji}), \quad \mbox{for $j\not=i$},\\
r_i^{n-1}e^{-\zeta r_i} y_{lm}(\hat{r}_i), \quad \mbox{for $j=i$},\end{cases}\\
\label{pco2} C'_{inlm,j\lambda\mu} & =  \begin{cases} 4\pi
\sum_{l''} I_R(lm,l'm',l''m'')
V_{ll'l''}^{n,p}(r,R_{ji})y_{l''m''}(\hat{R}_{ji}), \quad \mbox{for $j\not=i$},\\
(n-1)(r_i^{n-2}e^{-\zeta r_i} - r_i^{n-1}\zeta e^{-\zeta r_i})
y_{lm}(\hat{r}_i), \quad \mbox{for $j=i$},\end{cases}
\end{align}
where $I_R(lm,l'm',l''m'')$ is a Gaunt integral
$$ I_R(lm,l'm',l''m'') = \int y_{lm}y_{l'm'}y_{l''m''} dr^2,$$
$(R_{ji},\hat{R}_{ji})$ are the spherical coordinates for $X_i-X_j$,
and $V_{ll'l''}^n$ and $V_{ll'l''}^{n,p}$ are given by
\begin{align*}
V_{ll'l''}^n(r,R) & = (-1)^{n-1}
\frac{1}{\zeta^{n-1}}\sum_{i=0}^{l'}
\frac{\Gamma(l'+i+1)}{\Gamma(i+1)\Gamma(l'-i+1)}\sum_{j=0}^{l''}
\frac{\Gamma(l''+j+1)}{\Gamma(j+1)\Gamma(l''-j+1)}
\frac{1}{2^{i+j+1}}\\
& \times \sum_{k=0}^n
a_k^{nl}(l-i-j-1)\sum_{k'=0}^k\frac{\Gamma(k+1)}{\Gamma(k-k'+1)\Gamma(k'+1)}
(\zeta r)^{k-k'-i+1}(-1)^i\\
& \times [\exp(\zeta r) + (-1)^{l'+k-k'-i-1}\exp(-\zeta r)]
(-1)^{k'} (\zeta R)^{k'-j-1}\exp(-\zeta R), \end{align*}
\begin{align*}
V_{ll'l''}^{n,p}(r,R) & = (-1)^{n-1}
\frac{1}{\zeta^{n-1}}\sum_{i=0}^{l'}
\frac{\Gamma(l'+i+1)}{\Gamma(i+1)\Gamma(l'-i+1)}
\sum_{j=0}^{l''}\frac{\Gamma(l''+j+1)}{\Gamma(j+1)\Gamma(l''-j+1)}
\frac{1}{2^{i+j+1}}\\
&\times \sum_{k=0}^n
a_k^{nl}(l-i-j-1)\sum_{k'=0}^k\frac{\Gamma(k+1)}{\Gamma(k-k'+1)\Gamma(k'+1)}
\\
&\times \zeta^{k-k'-i+1}(-1)^{i+k'}(\zeta R)^{k'-j-1}\exp(-\zeta R)\\
&\times \big[ (k-k'-i-1)r^{k-k'-i-2}(\exp(\zeta r)
(-1)^{l'+k-k'-i-1}\exp(-\zeta r)) \\
& \quad + (\zeta r)^{k-k'-i-1} (\zeta\exp(\zeta r)
+(-1)^{l'+k-k'-i}\zeta\exp(-\zeta r))\big ],
\end{align*}
where $\Gamma$ is the Gamma function
$$\Gamma(n+1) = n!,$$
and $a_k^{np}(p)$ is defined by
$$\left(\frac{d}{dx}\right)^{n-1} \left(\frac{1}{x}
\frac{d}{dx}\right)^l x^p\exp(x) = \left(\sum_{k=0}^n
a_k^{nl}(p)x^k\right) x^{p-n-l}\exp(x).$$

\section{Numerical approximation and implementation}

In this section, we consider the discretizations for the real-space
LASTO method, its numerical approximations, and its implementation.

The Schr\"odinger equation is defined on an infinite domain and the
charge density is rapidly decaying and smooth at large distances
from the atoms. To accomplish an efficient discretization of the
infinite domain, we consider a large finite adaptive domain which
has fine meshes near the atoms and coarse meshes at large distances
from them. At the edge of this finite domain, we impose Dirichlet
boundary conditions.

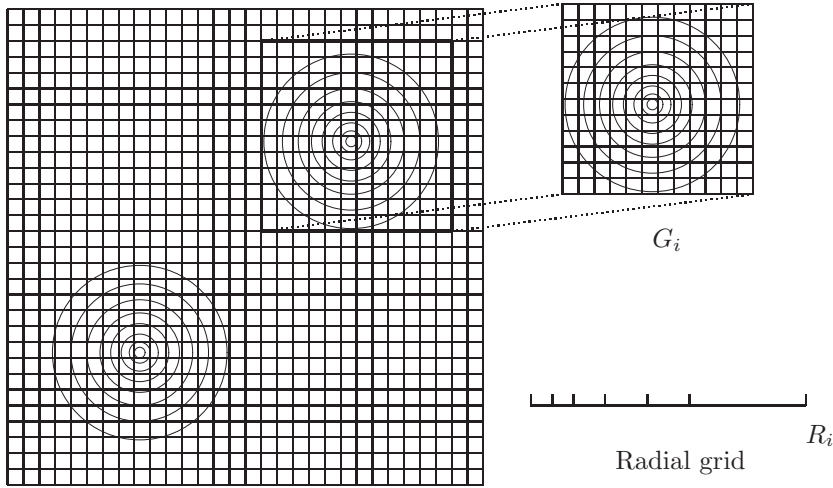
\begin{figure}
\begin{center}
\setlength{\unitlength}{2pt}
\begin{picture}(160,100)(0,0)
\put(30,30){\circle{2}}\put(30,30){\circle{4}}\put(30,30){\circle{7}}\put(30,30){\circle{11}}
\put(30,30){\circle{15}}\put(30,30){\circle{20}}
\put(30,30){\circle{26}}\put(30,30){\circle{33}}
\put(70,70){\circle{2}}\put(70,70){\circle{4}}\put(70,70){\circle{7}}\put(70,70){\circle{11}}
\put(70,70){\circle{15}}\put(70,70){\circle{20}}
\put(70,70){\circle{26}}\put(70,70){\circle{33}}
\put(5,5){\grid(90,90)(3,3)} \drawline(104,20)(156,20)
\drawline(104,20)(104,22)\drawline(108,20)(108,22)
\drawline(112,20)(112,22) \drawline(118,20)(118,22)
\drawline(126,20)(126,22) \drawline(134,20)(134,22)
\drawline(156,20)(156,22) \put(156,13){\makebox{$R_i$}}
\put(120,8){\makebox{Radial grid}}
\put(127,77){\circle{2}}\put(127,77){\circle{4}}\put(127,77){\circle{7}}
\put(127,77){\circle{11}}
\put(127,77){\circle{15}}\put(127,77){\circle{20}}
\put(127,77){\circle{26}}\put(127,77){\circle{33}}
\put(110,60){\grid(36,36)(3,3)} \put(127,50){\makebox{$G_i$}}
\dottedline{1}(110,60)(53,53) \dottedline{1}(110,96)(53,89)
\dottedline{1}(146,60)(89,53) \dottedline{1}(146,96)(89,89)
\thicklines \drawline(53,53)(53,89)(89,89)(89,53)(53,53)
\end{picture}
\caption{Discretization of domain showing overlay of atomic center
grids and background grid. A uniform mesh is used in the background grid while a logarithmic mesh is used inside the spheres.
}\label{discretization}
\end{center}\end{figure}

Because the basis functions are defined separately inside of the
spheres (muffin tin) and outside of the spheres (interstitial
region), we have to handle differently the regions interior and
exterior to the spheres and match the solutions on the sphere
boundaries. We use overset grids which consist of regular cubic grid
meshes in whole domain and exponential radial grid meshes inside the
spheres (Fig. \ref{discretization}).

Inside the spheres, we represent the electronic charge density and
the potential with linear combinations of real spherical harmonics,
i.e.,
\begin{equation}\label{lincom} \rho(\bold x) = \sum _{LM}
\rho_{LM}(r_i) y_{LM}(\hat{r}_i),\quad V(\bold x) = \sum _{LM}
V_{LM}(r_i) y_{LM}(\hat{r}_i).
\end{equation}

For computation, we have to restrict the $\lambda$ in (2.9) and the
maximum for $L$ in (2.14). We choose $8$ as a maximum for $\lambda$
in (2.9) and the maximum $4$ for $L$ in (3.1). These restrictions
affect the accuracy of the computations which are also affected by
the size of the finite domain, the mesh size of the regular grids
and the radial grids.

We next consider the implementation of the method. This 
consists of four steps: the computation of the potential, matrix
generation, the solution of the eigenvalue problem and updating the
charge density.

To obtain the potential, we solve the Poisson equation (2.6) and
(2.7) for the Coulomb potential in the interstitial region and
inside the spheres. Because of the definition of Coulomb potential as
the solution of the Poisson equation which has a source which
includes delta functions at the center of the atoms, we use a
pseudo-density $\tilde \rho$ to get the Coulomb potential in the
interstitial region at each grid point. We solve the Poisson
equation
\begin{align} \label{po2}
 \nabla^2 V_C(\bold x) &=  4\pi (\tilde\rho(\bold
x)),\\ V_C(\bold x) &= 0,\qquad \mbox{on the boundary}
\end{align}
with
\begin{align*} \int_0^{R_i} \tilde{\rho}(r) r^2 dr &= q =  \int_0^{R_i}
\left(\rho(r) - Z_i\delta_{C_i}\right) r^2 dr\\
\tilde{\rho}(R_i) & = \rho(R_i),\\
\frac{d \tilde{\rho}}{d r }(R_i) & = \frac{d \rho}{d r} (R_i).
\end{align*}
The pseudo-density $\tilde{\rho}$ has the same zeroth multipole
moments as $\rho$ as boundary values and the same derivative also on
the boundary of the spheres.

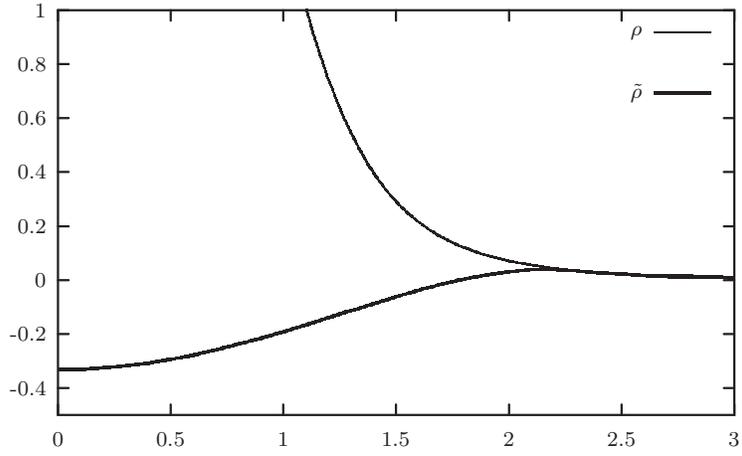
\begin{figure}
\begin{center}
\setlength{\unitlength}{0.2pt}
\begin{picture}(1500,900)(0,0)
\footnotesize \thicklines \path(154,141)(174,141) \thicklines
\path(1433,141)(1413,141) \put(132,141){\makebox(0,0)[r]{-0.4}}
\thicklines \path(154,243)(174,243) \thicklines
\path(1433,243)(1413,243) \put(132,243){\makebox(0,0)[r]{-0.2}}
\thicklines \path(154,345)(174,345) \thicklines
\path(1433,345)(1413,345) \put(132,345){\makebox(0,0)[r]{0}}
\thicklines \path(154,447)(174,447) \thicklines
\path(1433,447)(1413,447) \put(132,447){\makebox(0,0)[r]{0.2}}
\thicklines \path(154,550)(174,550) \thicklines
\path(1433,550)(1413,550) \put(132,550){\makebox(0,0)[r]{0.4}}
\thicklines \path(154,652)(174,652) \thicklines
\path(1433,652)(1413,652) \put(132,652){\makebox(0,0)[r]{0.6}}
\thicklines \path(154,754)(174,754) \thicklines
\path(1433,754)(1413,754) \put(132,754){\makebox(0,0)[r]{0.8}}
\thicklines \path(154,856)(174,856) \thicklines
\path(1433,856)(1413,856) \put(132,856){\makebox(0,0)[r]{1}}
\thicklines \path(154,90)(154,110) \thicklines
\path(154,856)(154,836) \put(154,45){\makebox(0,0){0}} \thicklines
\path(367,90)(367,110) \thicklines \path(367,856)(367,836)
\put(367,45){\makebox(0,0){0.5}} \thicklines \path(580,90)(580,110)
\thicklines \path(580,856)(580,836) \put(580,45){\makebox(0,0){1}}
\thicklines \path(793,90)(793,110) \thicklines
\path(793,856)(793,836) \put(793,45){\makebox(0,0){1.5}} \thicklines
\path(1007,90)(1007,110) \thicklines \path(1007,856)(1007,836)
\put(1007,45){\makebox(0,0){2}} \thicklines \path(1220,90)(1220,110)
\thicklines \path(1220,856)(1220,836)
\put(1220,45){\makebox(0,0){2.5}} \thicklines
\path(1433,90)(1433,110) \thicklines \path(1433,856)(1433,836)
\put(1433,45){\makebox(0,0){3}} \thicklines
\path(154,90)(1433,90)(1433,856)(154,856)(154,90)
\put(1259,814){\makebox(0,0)[r]{$\rho$}} \thinlines
\drawline(1281,814)(1389,814) \thinlines
\drawline(625,856)(637,813)(651,770)(665,728)(680,690)(695,653)(711,620)(727,589)(743,561)
(760,536)(777,513)(795,493)(814,474)(833,458)(852,444)(872,431)(893,420)(914,410)(936,402)
(958,394)(981,388)(1005,382)(1030,377)(1055,373)(1081,369)(1107,366)(1135,363)(1163,360)
(1192,358)(1222,356)(1252,355)(1284,354)(1316,352)(1350,351)(1384,350)(1419,350)(1433,349)
\put(1259,700){\makebox(0,0)[r]{$\tilde{\rho}$}}
\Thicklines\drawline(1281,700)(1389,700)
\Thicklines\drawline(154,176)(154,176)(197,176)(239,179)(282,183)(325,188)(367,195)(410,203)
(452,213)(495,224)(538,235)(580,247)(623,260)(666,274)(708,287)(751,300)(793,313)(836,326)
(879,337)(921,347)(964,355)(1007,361)(1049,365)(1092,365)(1135,363)(1177,359)(1220,357)
(1262,354)(1305,353)(1360,351)(1415,350)(1433,349)
\end{picture}
\caption{Full density $\rho$ and pseudodensity $\tilde \rho$ in units of $e/a_0^3$ versus distance in units of $a_0$ for $Pd$ atom.}
\label{pseudo1}
\end{center}\end{figure}

\begin{figure}
\begin{center}
\epsfig{file=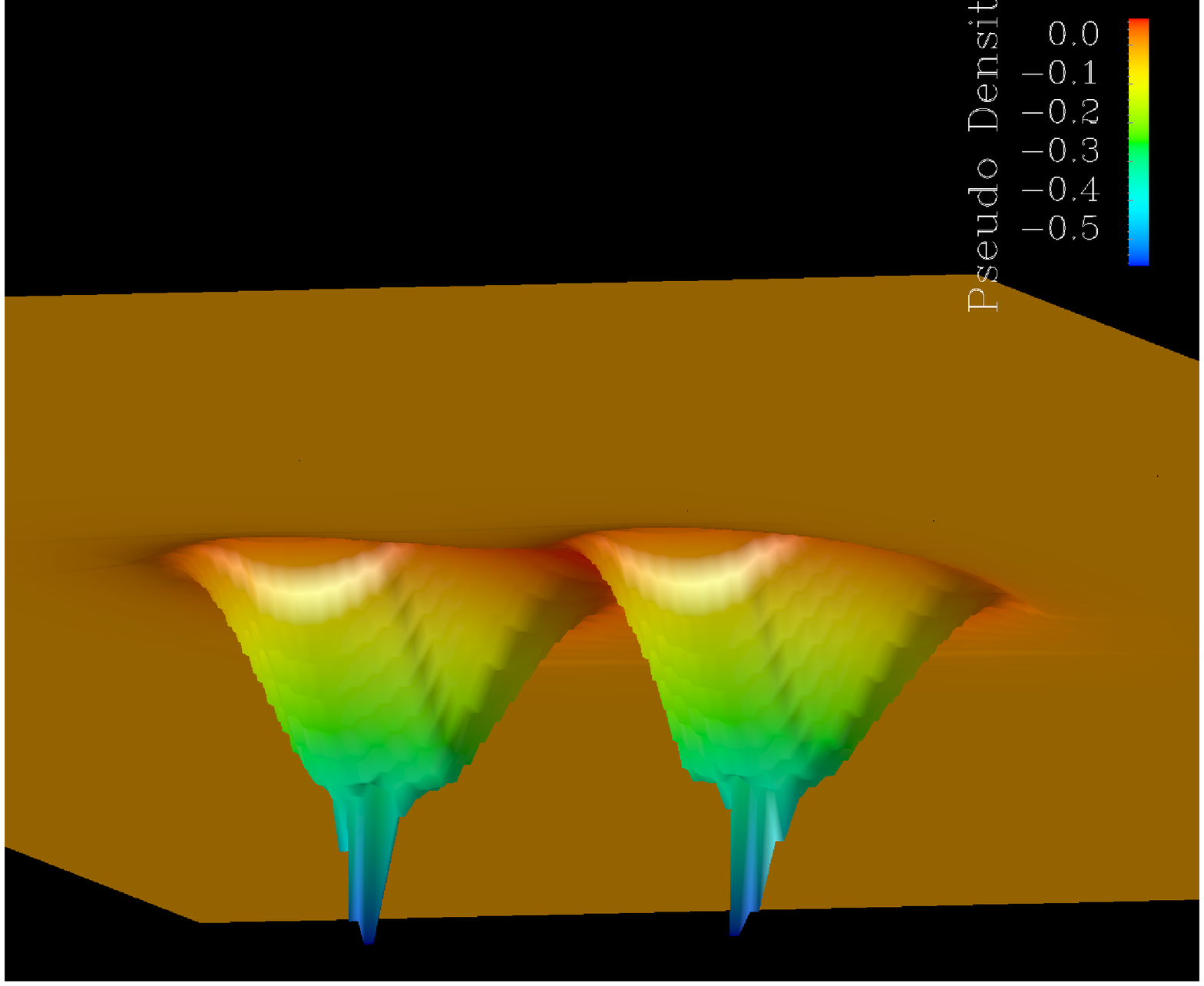 ,height=4.5cm}
\epsfig{file=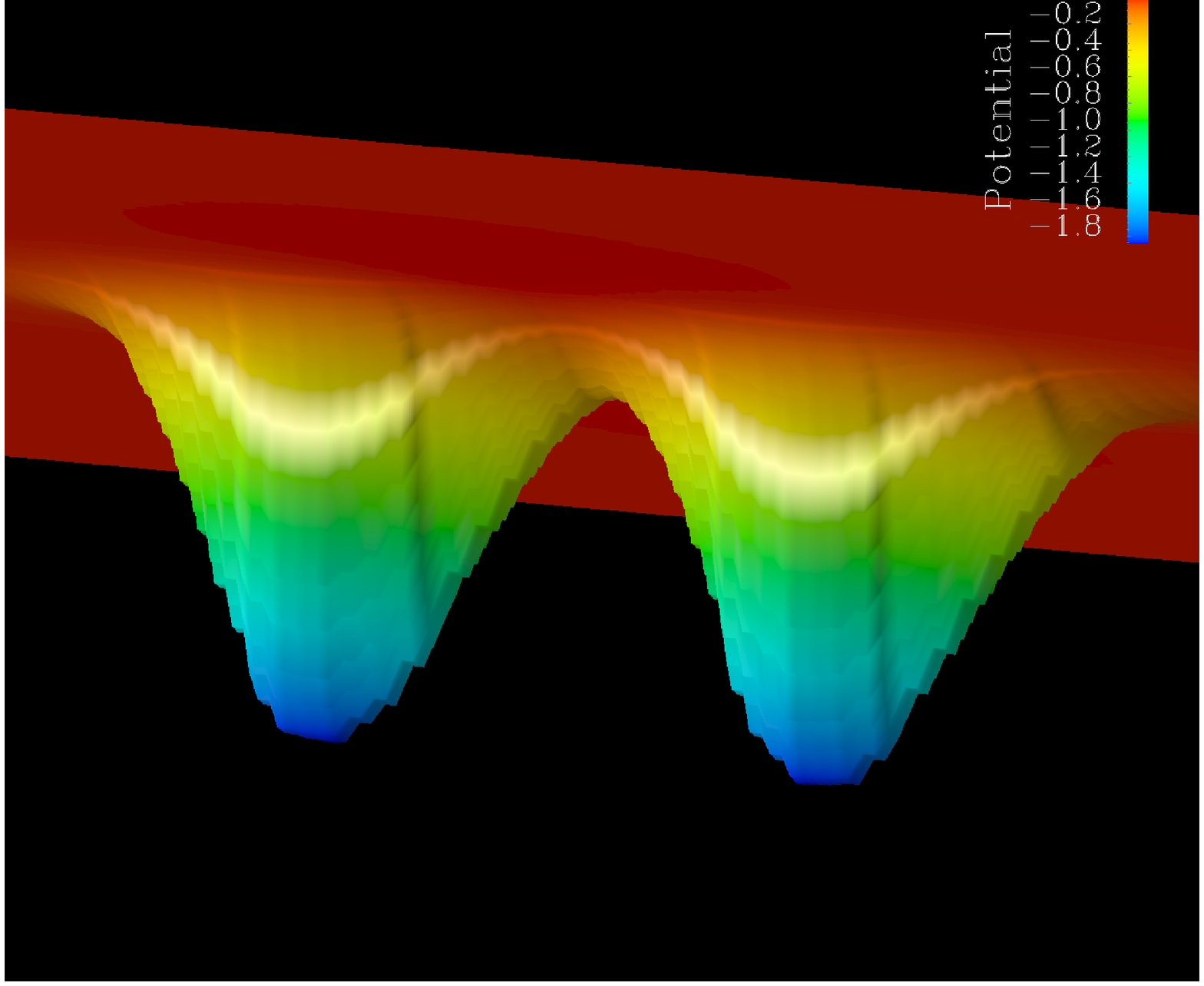 ,height=4.5cm}

(a) Pseudo-density \qquad\quad (b) Coulomb potential

\caption{The pseudo-density and Coulomb potential 
for a palladium dimer on the $z=0$ plane}\label{pseudo}
\end{center}\end{figure}

We plot the real density and pseudo-density in radial coordinates
for a single palladium atom in Fig. \ref{pseudo1} and the
pseudo-density and the solution of the Poisson equation for a
palladium dimer in Fig. \ref{pseudo}.

To solve the Poisson equation (\ref{po2}), we use a finite
difference scheme and a multigrid method \cite{bramble, brandt,
hackb,wesse}, which is a well-known, rapid, and scalable solver of
elliptic partial differential equations.

Then, we solve the spherical Poisson equation
\begin{equation}\label{sppo}
\left(\frac{\partial^2}{\partial r^2} +
\frac{2}{r}\frac{\partial}{\partial r} -
\frac{L(L+1)}{r^2}\right)V_{LM}^C(r_i) = 4\pi\left(\rho_{LM}(r_i) -
\delta_{L0}Z_i\right),
\end{equation} inside a sphere with Dirichlet boundary conditions which come
from the solution of (\ref{po2}), i.e.,
\begin{equation}\label{poie2} V_{\mbox{\scriptsize
C}}(\bold{x}) = \sum_{LM} V_{LM}^C(r_i) y_{LM}(\hat{r}_i)
\end{equation} where
\begin{align*}
V_{LM}^C(r_i) &= V_{LM}^C(R_i)\left(\frac{r_i}{R_i}\right)^L +
\frac{4\pi}{2L+1}
\left(\frac{1}{r_i^{L+1}}\int_0^{r_i}\rho_{LM}(r')^{L+2}dr'\right. \\
-\frac{Z_i}{\sqrt{4\pi}}&
\left(\frac1{r_i}-\frac1{R_i}\right)\delta_{L0} \left.
-\frac{r_i^L}{R_i^{2L+1}}\int_0^{R_i}\rho_{LM}(r')^{L+2} dr'
+r_i^L\int_{r_i}^{R_i}\rho_{LM}(r')^{L-l}dr'\right)
\end{align*}
and
\begin{align}\label{poie3} V_{LM}^C(R_i) & = \int_{\partial S_i}
V_{\mbox{\scriptsize C}}(\bold x) y_{LM}(\bold x)
dS\\
\label{poie4} \rho_{LM}(R_i) &= \int_{\partial S_i} \rho (\bold x)
y_{LM}(\bold x) dS.\end{align}

Next, we consider the generation of the overlap matrix $S$ and the
Hamiltonian matrix $H$ in the interstitial region and the spheres.
From (2.1) we obtain
\begin{equation}\label{weakform}
\int_{\Omega} \varphi_{i}(\bold x)\left\{ -\frac12\nabla^2 + V(\bold
x)\right\}\varphi_{k}(\bold x)d\bold x = \int_{\Omega}
\varphi_{i}(\bold x)\varphi_{k}(\bold x)d\bold x.
\end{equation}
Using Green's theorem the left-hand side (the Hamiltonian matrix) of (\ref{weakform}) can
be rewritten
\begin{equation*}\begin{split}
\int_{\Omega} \varphi_{i}(\bold x)& \left\{ -\frac12\nabla^2 +
V(\bold x)\right\}\varphi_{k}(\bold x)d\bold x \\
& = \int_{\Omega} \left(\frac12 \nabla \varphi_{i}(\bold x) \nabla
\varphi_{k}(\bold x) + \varphi_i(\bold x) V(\bold
x)\varphi_{k}(\bold x)\right) d\bold x -\frac12 \int_{\partial
\Omega} \nabla \varphi_i \frac{\partial \varphi_k}{\partial \bold n}
d\bold x
\end{split}
\end{equation*}
Thus, the Hamiltonian matrix is symmetric with
zero Dirichlet or Neumann boundary conditions, but not in general.
Thus we consider the following symmetric form derived from Green's
theorem and zero Dirichlet or Neumann boundary condition,
\begin{equation}\label{weakori}
\int_{\Omega} \left(\frac12 \nabla \varphi_{i}(\bold x) \nabla
\varphi_{k}(\bold x) + \varphi_i(\bold x) V(\bold
x)\varphi_{k}(\bold x)\right) d\bold x = \int_{\Omega}
\varphi_{i}(\bold x)\varphi_{k}(\bold x)d\bold x.
\end{equation}

The evaluation of (\ref{weakori}) requires considerable
computational effort, so we use the following symmetrized form
\begin{equation}\label{weaksym}
\begin{split}
\frac12 \int_{\Omega} &\varphi_{i}(\bold x)\left\{ -\frac12\nabla^2
\right\}\varphi_{k}(\bold x) + \varphi_{k}(\bold x) \left\{
-\frac12\nabla^2 \right\}\varphi_{i}(\bold x)d\bold x  \\
&+ \int_{\Omega} \varphi_i(\bold x) V(\bold x)\varphi_{k}(\bold x)
d\bold x  = \int_{\Omega} \varphi_{i}(\bold x)\varphi_{k}(\bold
x)d\bold x,
\end{split}
\end{equation}

In the interstitial region, we compute the contribution  of the
matrices  $S^*$ and $H^*$ by
\begin{align}\label{poie4.1}
\begin{split}H_{inlm,i'n'l'm'} &= \frac12 \int_{\Omega-\cup_i S_i} \phi_{inlm}(\bold
x)\left\{ -\frac12\nabla^2 + V(\bold
x)\right\}\phi_{i'n'l'm'}(\bold x)\\
& + \phi_{i'n'l'm'}(\bold x)\left\{ -\frac12\nabla^2 + V(\bold x)
\right\}\phi_{inlm}(\bold x)d\bold x
\end{split}\\
\label{poie4.2} S_{inlm,i'n'l'm'} &= \int_{\Omega-\cup_i S_i}
\phi_{inlm}(\bold x)\phi_{i'n'l'm'}(\bold x)d\bold x,
\end{align}
where \begin{equation}\label{poie4.3} \phi_{inlm}(\bold x) =
r_i^{n-1}\exp(-\zeta r_i)y_{lm}(\hat{r}_i).
\end{equation}

To compute the contribution of the matrices $S^*$ and $H^*$, we use
the grid or mid-point of each cube, i.e.,
$$
S^1_{inlm,i'n'l'm'} = \sum_{P} \phi_{inlm}(P)\phi_{i'n'l'm'}(C_P)
\mbox{Vol}(C_P),$$ or
\begin{align*}
S^2_{inlm,i'n'l'm'} &=
\sum_{C}\int_{C}\phi_{inlm}(\bold x)\phi_{i'n'l'm'}(\bold x)d\bold x \\
&= \sum_{C}\phi_{inlm}(C)\phi_{i'n'l'm'}(C)\mbox{Vol}(C_C),
\end{align*}
where $P$'s are the points of a uniform grid, $C$'s are the center
of cube in uniform grid, and $\mbox{Vol}(C_P)$ is the volume of a
cube centered $P$. In each case, we have to compute $\phi_{ilmn}$ on
all grid points for each $i,n,l,m$.

For the spherical coordinate operator
$$ \nabla^2 = \frac{\partial ^2}{\partial r^2}
+\frac{2}{r}\frac{\partial}{\partial r} + \frac{1}{r^2} \left[
\frac{1}{\sin^2\phi}\frac{\partial^2 }{\partial \theta^2}
+\frac{\cos \phi}{\sin \phi}\frac{\partial }{\partial \phi}
+\frac{\partial ^2}{\partial \phi^2}\right]$$ and
$$\left[
\frac{1}{\sin^2\phi}\frac{\partial^2 }{\partial \theta^2}
+\frac{\cos \phi}{\sin \phi}\frac{\partial }{\partial \phi}
+\frac{\partial ^2}{\partial \phi^2} + l(l+1)\right] y_{lm} = 0,$$
we have
\begin{equation}\label{poie4.4}
\begin{split} \ & \left[-\frac12\nabla^2 + V(\bold x)\right]
\phi_{inlm}(\bold x)\\
 &= \frac12\left[(l(l+1)-n(n-1))r_i^{n-3} +2\zeta n
r_i^{n-2} -\zeta^2 r_i^{n-1}\right]\exp(-\zeta r_i) y_{lm}(\hat{r}_i)\\
&+ V(\bold x) r_i^{n-1}\exp(-\zeta r_i)y_{lm}(\hat{r}_i) =
\tilde{\phi}_{inlm}(\bold x).
\end{split}
\end{equation}
So, we have to compute
\begin{align*}
H^1_{inlm,i'n'l'm'} &= \frac12\sum_{P}\left(
\phi_{inlm}(P)\tilde{\phi}_{i'n'l'm'}(P) + \phi_{i'n'l'm'}(P)\tilde{\phi}_{inlm}(P) \right)\mbox{Vol}(P),\\
S^1_{inlm,i'n'l'm'} &= \sum_{P} \phi_{inlm}(P)\phi_{i'n'l'm'}(P)
\mbox{Vol}(P),
\end{align*}
i.e., $\phi_{inlm}(P)$ and $\tilde{\phi}_{inlm}(P)$ at each regular
cubic grid point which does not include any spheres.

For the contribution of $S$ and $H$ inside spheres, we can compute
the elements of the matrices with given $\beta$'s and $\alpha$'s as
in \cite{daven}.
\begin{align*}
S_{inlm,i'l'm'n'}^j & = \int_{S_j}
\phi_{inlm}(\bold{x})\phi_{i'l'm'n'}(\bold{x}) d\bold{x}\\
& = \sum_{j\lambda\mu}(\beta_{inlm,j\lambda\mu}
\beta_{i'n'l'm',j\lambda\mu} + \alpha_{inlm,j\lambda\mu}
\alpha_{i'n'l'm',j\lambda\mu}<\dot{g}_{j\lambda}|
\dot{g}_{j\lambda}>),\\
H_{inlm,i'l'm'n'}^j & = \frac12\int_{S_j}\left(
\phi_{inlm}(\bold{x})\left\{-\frac12\nabla^2 + V(\bold{x})\right\} \phi_{i'l'm'n'}(\bold{x}) \right.\\
&\left. \phi_{i'n'l'm'}(\bold{x})\left\{-\frac12\nabla^2 +
V(\bold{x})\right\}  \phi_{ilmn}(\bold{x})
\right)d\bold{x}\\
& = e_jS_{inlm,i'l'm'n'}^j + \frac12\sum_{j\lambda\mu}\left(
\beta_{inlm,j\lambda\mu}\alpha_{i'n'l'm',j\lambda\mu} +
\alpha_{inlm,j\lambda\mu}\beta_{i'n'l'm',j\lambda\mu}\right)\\
&+ \sum_{\lambda\mu,\lambda'\mu',LM,L>0} \int_0^{R_j}
[\beta_{inlm,j\lambda\mu} g_{j\lambda}(r) +
\alpha_{inlm,j\lambda\mu} \dot{g}_{j\lambda}(r)]V_{LM}(r)\\
& \cdot [\beta_{i'n'l'm',j\lambda'\mu'} g_{j\lambda'}(r) +
\alpha_{i'n'l'm',j\lambda'\mu'} \dot{g}_{j\lambda'}(r)]r^2 dr
I_R(\lambda\mu,\lambda'\mu',LM),
\end{align*}
where $I_R(\lambda\mu,\lambda'\mu',LM)$ is a real Gaunt Integral.

 To solve the eigenvalue problems $H\varphi_k = \epsilon_k
S\varphi_k$ where $H$ and $S$ are real symmetric and dense matrices
and $S$ is positive definite with dimension $n =
N_{\mbox{\scriptsize basis}}$, we use ScaLAPACK \cite{scalapack},
which is well developed and optimized for parallel machines.

%\begin{rem}
%\end{rem}

Here, we consider updating the charge density from the
solutions of $H\varphi_k = \epsilon_k S\varphi_k$.

We have to compute $\rho_{\mbox{\scriptsize out}} =
\sum_{\epsilon_k<E_{\mbox{\scriptsize Fermi}}} |\varphi_k|^2
+\rho_{\mbox{\scriptsize core}}$ in the whole region including both
the spheres and the interstitial region. We write
$\varphi_k$
$$\varphi_k = \sum_{inlm} \varphi_{k,inlm}\phi_{inlm}.$$

In the interstitial region, we compute
$$\rho_{\mbox{\scriptsize out}}(\bold x) = \sum_{\epsilon_k < E_{\mbox{\scriptsize Fermi}}}
|\varphi_k(\bold x)|^2, $$
where
$$\varphi_k(\bold x) = \sum_{inlm} \varphi_{k,inlm}(r_i^{n-1}e^{-\zeta
r_i}y_{lm}(\hat{r}_i))$$ at all regular cubic grid points $\bold x$.

Inside the spheres, we compute
\begin{equation} \label{update}\rho(\bold x)
=\sum_{jLM} \rho_{jLM} y_{LM}(\hat{r}_j) +\rho_{\mbox{\scriptsize
core}} = \sum_{\epsilon_k < E_{\mbox{\scriptsize Fermi}}}
|\varphi_k(\bold x)|^2 + \rho_{\mbox{\scriptsize core}}.
\end{equation}
From the orthogonality of $g$ and $\dot{g}$, we have
\begin{align*}\rho_{jLM} &= \int_{\partial S_j(r_j)} \varphi_k^2
y_{LM}(\hat{r}_j) ds \\
&=\sum_{inlm}\sum_{i'n'l'm'} \sum_{j\lambda\mu}\sum_{j\lambda'\mu'}
\varphi_{k,inlm}\varphi_{k,i'n'l'm'}[\beta_{inlm,j\lambda\mu}g_{j\lambda}(r_j)
+\alpha_{inlm,j\lambda\mu}\dot{g}_{j\lambda\mu}(r_j)]\\
&[\beta_{i'n'l'm',j\lambda'\mu'}g_{j\lambda'}(r_j)
+\alpha_{i'n'l'm',j\lambda'\mu'}\dot{g}_{j\lambda'\mu'}(r_j)]
I_R(LM,\lambda\mu,\lambda'\mu').
\end{align*}

\section{Results}

To test the method we have calculated the total energy for several palladium clusters.

The total energy is computed in the usual way
\begin{align*}
E & = -\sum_k^{\mbox{occ}} \int_{\Omega} \varphi_k(\bold x)
\frac{\nabla^2}{2} \varphi_k(\bold x) d \bold x + \int_\Omega
V_{\mbox{\scriptsize Ext}}(\bold x) \rho(\bold x) d\bold x\\
& + \frac 12 \int_\Omega \int_\Omega \frac{\rho(\bold x) \rho(\bold
x')}{|\bold x-\bold x'|} d\bold x' d\bold x + E_{\mbox{\scriptsize
xc}},
\end{align*}
where the terms are respectively the non-interacting kinetic energy,
the external potential, the Hartree and the exchange and correlation
energies. This formula can be simplified using (\ref{sch1}) to yield
\begin{equation}\label{energy}
E = \sum_{k}^{\mbox{occ}} \epsilon_k - \frac12 \int_\Omega
V_{\mbox{\scriptsize Hartree}}(\bold x)\rho(\bold x) d \bold x -
\int_\Omega V_{\mbox{\scriptsize xc}}(\bold x)\rho(\bold x) d \bold
x + E_{\mbox{\scriptsize xc}},
\end{equation}
where
$$E_{\mbox{\scriptsize xc}} = \int_\Omega e_{\mbox{\scriptsize xc}}(\bold x)\rho(\bold x) d \bold
x.$$

There have been many calculations of the energetics of Pd clusters in recent years \cite{Moseler, Kumar, zhang, Chang, Longo}. Most calculations show that small clusters are magnetic (spin polarized) and for example that $Pd_{13}$ has an icosohedral or possibly a buckled biplanar structure \cite{Kumar}. Since our purpose is to test the method, we have compared results for non-magnetic structures in simpler geometries. 

The LASTO basis set was optimized for crystalline palladium in the facecentered cubic structure. It consistes of two $s$, two $p$, and two $d$ functions plus one $f$ function per atom. The $\zeta$ values are given in Table \ref{basis}.

\begin{table}
\centering \caption{LASTO Basis and $\zeta$ values} \label{basis}
\begin{tabular}{|c|c|}
\hline  function & $\zeta(a_0^{-1})$ \\ \hline
4d & 2.20\\ \hline
5d & 2.60\\ \hline
5s & 1.00\\ \hline
5p & 1.00\\ \hline 
6s & 2.00\\ \hline
6p & 2.00\\ \hline
4f & 1.60\\ \hline
\end{tabular}
\end{table}

The 1s through 4p states were treated as core functions. They are solved using the full Dirac equation in the spherical part of the potential within the spheres. 
A logarithmic radial mesh was used inside the spheres, with 429 points. The (uniform) mesh in the interstitial region had a spacing of $0.05$ $a_0$. Calculations with NWchem used the SBKJC basis with an effective core potential given in \cite{Stevens}. In both sets of calculations states near the highest occupied level were smeared with a Gaussian of $\approx 0.001 au$.

\begin{figure}
\begin{center}
\epsfig{file=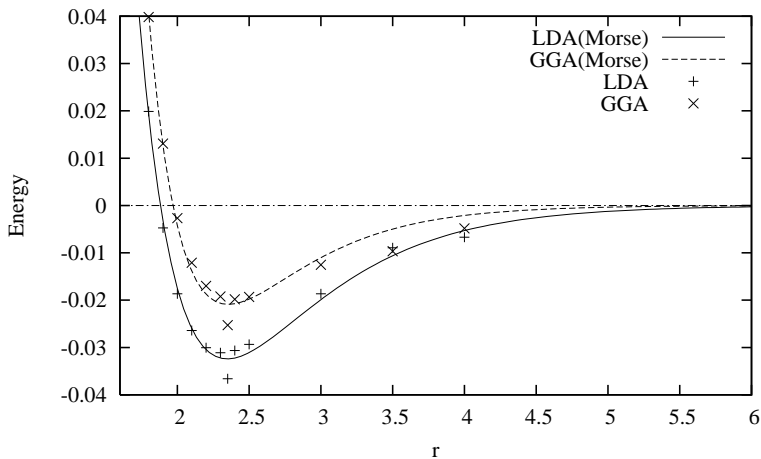 ,height=6cm} \caption{Binding energy of
Pd$_{2}$ in units of Hatrees/atom as a function of the separation distance
$r$ in atomic units. The smooth curves are fits to a Morse potential.}\label{pd2bind}
\label{dimer}
\end{center}
\end{figure}

In Fig. \ref{dimer} we plot the binding energy of a palladium dimer as a function of $r$ with LSD and GGA expressions for exchange and correlation energy. Here we computed the binding energy per atom by
\[
E = \frac{E(Pd_n) -nE(Pd)}{n}
\]
and fitted
the data with Morse potential energy function, i.e.,
$$ E(r) = D\left\{ e^{-2a(r-R_0)} - 2e^{-a(r-R_0)}\right\}$$
where D is the minimum energy, $R_0$ is the equilibrium internuclear distance.

In Table \ref{pd2c}, we compare the results with experiment \cite{Lin}\cite{Shim}\cite{huber} and
with numerical results of Zhang, Ge, and Wang \cite{zhang}, which
use plane wave basis sets and GGA for exchange and correlation
energy, and DMol$^3$ \cite{dmol3} program which uses LCAO (Linear
combination of atomic orbital) with several methods for the exchange
and correlation energy. Our results for both the internuclear distance and the binding energy are in agreement with previous calculations.

\begin{table}
\centering \caption{Binding energy for a Palladium (Pd) dimer with
comparing previous results} \label{pd2c}

\begin{tabular}{|c|c|c|}
\hline  & binding energy & $R_0$ \\ \hline
Experiment \cite{Lin} & $0.37 \pm 0.13$ eV/atom& 4.71 au\\ \hline
Experiment \cite{Shim} & $0.52 \pm 0.08$ eV/atom & 4.71 au \\ \hline
Experiment \cite{huber}& 0.698 eV/atom & 4.71 au\\ \hline
 LASTO(PBE-GGA) & 0.559 eV/atom & 4.802 au \\ \hline
LASTO(LSD) &  0.882 eV/atom & 4.604 au\\
\hline
Plane wave, GGA \cite{zhang}& 0.63 eV/atom & 4.70 au\\
\hline DMol$^3$(PBE-GGA -- LCAO)\cite{dmol3} & 0.415 eV/atom & 4.70 au\\
\hline
 DMol$^3$(PW91-GGA -- LCAO) \cite{dmol3}& 0.538 eV/atom & 4.70 au\\ \hline
DMol$^3$(PW-LSD -- LCAO)\cite{dmol3} & 0.770 eV/atom & 4.70 au\\
\hline
NWChem (PBE-GGA)\cite{nwchem} & 0.468 eV/atom & 4.802 au\\
\hline
\end{tabular}
\end{table}

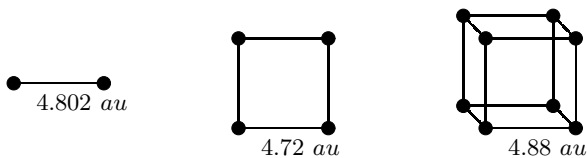
\begin{figure}
\begin{center}
\setlength{\unitlength}{1.7pt}
\begin{picture}(150,60)(0,20)
\put(10,50){\circle*{3}} \put(30,50){\circle*{3}}
\drawline(10,50)(30,50) \put(15,44){\mbox{\small 4.802 $au$}}
\put(60,60){\circle*{3}} \put(80,60){\circle*{3}}
\put(60,40){\circle*{3}} \put(80,40){\circle*{3}}
\drawline(60,60)(80,60)(80,40)(60,40)(60,60)
\put(65,34){\mbox{\small 4.72 $au$}} \put(110,65){\circle*{3}}
\put(130,65){\circle*{3}} \put(110,45){\circle*{3}}
\put(130,45){\circle*{3}} \put(115,60){\circle*{3}}
\put(135,60){\circle*{3}} \put(135,40){\circle*{3}}
\put(115,40){\circle*{3}}
\drawline(110,65)(130,65)(130,45)(110,45)(110,65)(115,60)(115,40)(110,45)
\drawline(115,40)(135,40)(130,45)
\drawline(135,40)(135,60)(115,60)(135,60)(130,65)
\put(120,34){\mbox{\small 4.88 $au$}}
\end{picture} \caption{Structures of dimer and 4- and 8- Pd
clusters used to test the LASTO code}\label{pdpic}
\end{center}\end{figure}

In Fig. \ref{pdpic}, we present the structures of Pd$_2$, Pd$_4$,
and Pd$_8$. We calculate the binding energy for Pd$_2$, for Pd$_4$,
and for Pd$_8$ with LSD and GGA methods for the exchange and
correlation energy. The bond lengths were chosen to be the same as \cite{zhang} and were not varied. In Table \ref{pdtable}, we present the results and compare with the results of \cite{zhang} and with \cite{Futschek}. Again the energies show agreement with previous work.

\begin{table}
\centering \caption{Binding energy in $eV/atom$ for Pd$_2$, Pd$_4$, and Pd$_8$
}\label{pdtable}
\begin{tabular}{|c|c|c|c|}
\hline  & Pd$_2$ & Pd$_4$ & Pd$_8$\\ 
\hline LASTO (LSD) &  0.861   & 1.633  & 2.700 \\
\hline LASTO (PBE-GGA) &  0.559   & 1.265  & 2.189 \\
\hline Plane wave (GGA) & .473 & 1.234 & 1.995 \\
\hline Plane wave (GGA) \cite{zhang} & 0.63   &  1.46 & 1.91 \\
\hline NWChem (PBE-GGA) &  0.468   & 1.405  & 1.869 \\
\hline\end{tabular}
\end{table}

\section{Conclusion}

In summary we have shown that a real space version of the linear augmented Slater type orbital method provides good agreement with other calculations for small clusters. The full treatment of core orbitals while maintaining a small basis set will be useful for treating heavy elements such as Pt or Au which are important nanoscale systems.

\section*{Acknowledgements}
We acknowledge assistance from Dimitri Volja and Jin-Cheng Zheng in programming the Slater type orbital expansion coefficients, 
eq. 2.12. We thank S. Chaudhuri for kindly providing results for the Pd dimer using $DMol^3$ \cite{dmol3}. This manuscript has been authored in part by Brookhaven Science Associates, LLC, under Contract No. DE-AC02-98CH10886 with the U.S. Department of Energy.

\end{document}